\documentclass[10pt, a4paper, twocolumn, aps, pra, showpacs, longbibliography, superscriptaddress]{revtex4-1}
\pdfoutput=1
\usepackage[utf8x]{inputenc}
\usepackage{ucs}
\usepackage{amsmath}
\usepackage{amsfonts}
\usepackage{amssymb}
\usepackage{makeidx}
\usepackage{cellspace,booktabs}
\usepackage{natbib}
\usepackage{lipsum}
\usepackage{soul}

\usepackage{color}
\definecolor{light-gray}{gray}{0.55}

\usepackage{microtype}

\usepackage{graphicx}

\renewcommand{\dag}{^{\dagger}}
\newcommand{\exv}[1]{ \langle #1 \rangle }

\newcommand{\bra}[1]{ \langle #1 \rvert }
\newcommand{\ket}[1]{ \lvert #1 \rangle}
\newcommand{\braket}[2]{\langle #1 \vert #2 \rangle }

\newcommand{\overlr}[1]{\overset{\leftrightarrows}{#1}}
\renewcommand{\overleftarrow}[1]{\overset{\leftarrow}{#1}}
\renewcommand{\overrightarrow}[1]{\overset{\rightarrow}{#1}}

\newcounter{appcounter}
\newenvironment{appendixalign}{%
\addtocounter{equation}{-1}
\refstepcounter{appcounter}

\begin{equation}}
{\end{equation}}

\begin{document}

\begin{abstract}
We propose to use a nonlinear resonator for projective readout, classical memory, and feedback for a superconducting qubit. Keeping the classical controller at cryogenic temperatures sidesteps many of the inefficiencies inherent in two-way communication between temperature stages in typical systems with room temperature controllers, and avoids increasing the cryogenic heat load.  This controller may find a broad range of uses in multi-qubit systems, but here we analyze two specific demonstrative cases in single qubit-control. In the first case, the nonlinear controller is used to initialize the qubit in a definite eigenstate.  And in the second case, the qubit's state is read into the controller's classical memory and used to reinstate the measured state after the qubit has decayed. We analyze the properties of this system and we show simulations of the time evolution for the full system dynamics.
\end{abstract}

\date{\today}
\author{Christian Kraglund Andersen}
\thanks{E-mail: ctc@phys.au.dk}
\affiliation{Department of Physics and Astronomy, Aarhus University, DK-8000 Aarhus C, Denmark}
\author{Joseph Kerckhoff}
\altaffiliation{Current address: HRL Laboratories, LLC, Malibu, CA 90265, USA}
\affiliation{JILA, University of Colorado, Boulder, Colorado 80309, USA}
\author{Konrad W. Lehnert}
\affiliation{JILA, University of Colorado, Boulder, Colorado 80309, USA}
\affiliation{National Institute of Standards and Technology, Boulder, Colorado 80305, USA}
\author{Benjamin J. Chapman}
\affiliation{JILA, University of Colorado, Boulder, Colorado 80309, USA}
\author{Klaus Mølmer}
\affiliation{Department of Physics and Astronomy, Aarhus University, DK-8000 Aarhus C, Denmark}

\title{Closing a quantum feedback loop inside a cryostat: Autonomous state-preparation and long-time memory of a superconducting qubit}

\maketitle

\section{Introduction}

The field of circuit QED \cite{wallraff2004strong,blais2004cavity}, which integrates superconducting resonators and qubits, has recently become a promising candidate for quantum information processing \cite{ chow2014implementing, barends2014superconducting, corcoles2015demonstration} due to dramatic improvements in the quantum technologies \cite{PhysRevLett.107.240501, PhysRevLett.111.080502,  PhysRevB.86.100506}. Coherent operations can be performed fast and with high fidelities \cite{barends2014superconducting, PhysRevA.87.030301}, and different strategies for state initialization \cite{johnson2012heralded, riste2012initialization, geerlings2013demonstrating, campagne2013persistent} and readout \cite{riste2012feedback, mallet2009single, PhysRevLett.112.190504, reed2010high} have been demonstrated.  However, the pursuit of large-scale quantum information processing is necessarily accompanied by a massive, further increase in quantum and classical system complexity for initialization, measurement, and control.  As complexity increases, the inefficiencies inherent in the standard approach to measurement and control using external, room temperature apparatuses will become increasingly burdensome.  Currently, circuit QED components are operated below \mbox{100 mK}, and measurement signals used for readout and conditional control are sent through an expensive and meter-scale amplification chain up to room temperature. Similarly, controlling the quantum components requires transmitting signals from room temperature through $\sim$60 dB of cold attenuation to the \mbox{${<} 100$ mK}-stage \cite{corcoles2015demonstration}.  Thus, the current approach to inter-temperature, two-way communication between quantum and classical hardware involves inefficiencies in size, power, cost and speed that may prove incompatible with quantum information processing beyond the few-qubit regime.

\begin{figure}[t]
\includegraphics[width=\linewidth]{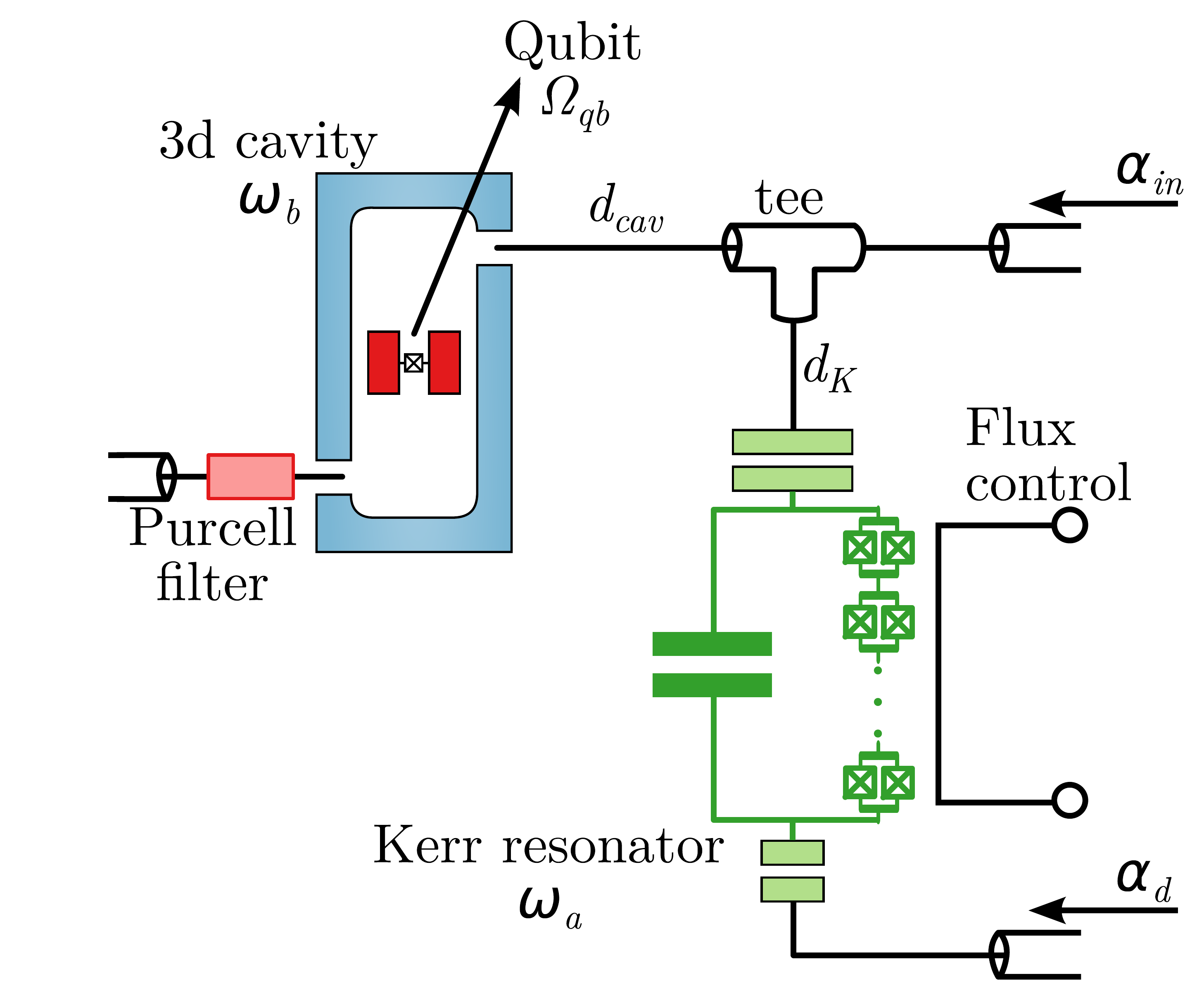}
\caption{(Color online) A qubit-cavity system is connected to a Kerr-resonator through a tee-junction. The qubit with a transition frequency $\Omega_{qb}$ is embedded in the 3d-cavity with a resonance at $\omega_b$ and the coupling port for the 3d-cavity is placed a distance $d_{cav}$ from the tee-junction. The Kerr-resonator, which consists of an array of SQUIDs that allows for a tunable frequency $\omega_a$, is placed a distance $d_K$ from the tee-junction. The Kerr-resonator is driven by a strong drive $\alpha_d$, while both the Kerr-resonator and the 3d-cavity are driven by the field $\alpha_{in}$. Both drives are operated at the same frequency $\omega_d$. } \label{fig:circuit}
\end{figure}

Another possible approach that sidesteps these inefficiencies would be to perform much of the measurement and actuation at the same temperature and with the same technology as the quantum hardware.  In this article we analyze a specific case of this alternate approach in which a nonlinear, superconducting microwave resonator \cite{castellanos2008amplification} effectively reads-out and feeds back to a superconducting 3d-qubit \cite{PhysRevLett.107.240501}. The nonlinear resonator acts like a Josephson bifurcation amplifier (JBA) \cite{mallet2009single, vijay2009invited, schmitt2014multiplexed} in that the circuit's Kerr-nonlinearity is driven and latches into one of two bistable states \cite{PhysRevLett.109.153602}, depending on the state of the qubit. In this way, the JBA implements a quantum non demolition (QND) readout of the qubit state and stores this state in its one-bit classical memory. Moreover, this classical memory-state may be maintained even if the qubit subsequently decays.  However, the Kerr-resonator does not play the standard JBA role, as communicating this state to the room temperature equipment (and experimenters) is not critical.  Rather, a state stored in the Kerr-resonator is fed back directly to implement conditional qubit operations.

Although we anticipate that natural advantages may be most practical in multi-qubit systems, here we analyze two, specific demonstrative cases in single qubit-control.  In the first case, the nonlinear controller is used to initialize the qubit in a definite eigenstate.  And in the second case, the qubit's state is read into the controller's classical memory, where it is stored for an indefinite period of time, and then used to reinstate the measured state after the qubit has decayed.  The outline of of the article is as follows: In Sec. II, we introduce the system components and formalise the protocols for the two schemes. In Sec. III, we recall the basic properties of the Josephson bifurcation amplifier, which is a central component in the proposed microwave network and we also present our theoretical formalism for the fully coupled system. In Sec. IV, we present numerical simulations with experimentally realistic component parameters and in Sec. V we compare our scheme with other approaches for feedback and control. Sec. VI concludes the article and presents an outlook.

\section{Qualitative description of the scheme}
\label{sec:Scheme}

The proposed setup is sketched in Fig. \ref{fig:circuit}. The two critical components are a two-port Kerr-resonator \cite{castellanos2008amplification} and a 3d-cavity containing a superconducting qubit \cite{PhysRevLett.107.240501}. These are interconnected via two transmission lines and a tee-junction.  The microwave network has three connections to the external environment.  The two most important are the transmission lines emanating from the bottom, weakly coupled port of the Kerr-resonator, and from the tee-port. A less critical port from the 3d-cavity uses a Purcell filter \cite{PhysRevLett.112.190504, reed2010purcell, bronn2015reducing, sete2015quantum} to control the photon number in the 3d-cavity and helps optimize performance. We will only discuss this port briefly in this section as the essential scheme can be understood without it. Dynamical control of the network is achieved by two microwave drives, labelled by the incoming amplitudes $\alpha_{d}$ and $\alpha_{in}$, and a flux control line that controls the center frequency of the Kerr-resonator.

\begin{figure}[b]
\includegraphics[width=\linewidth]{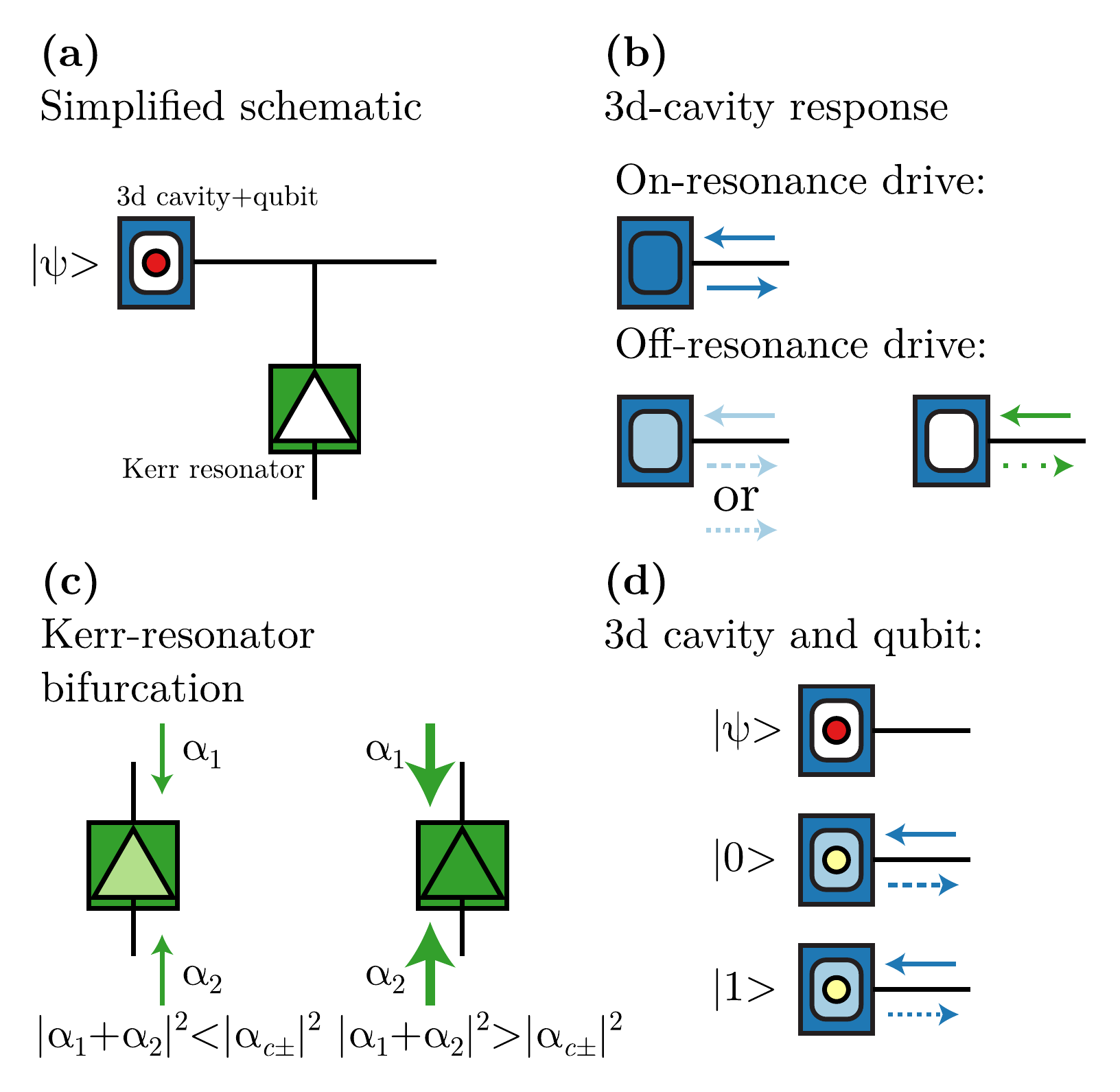}
\caption{(Color online) (a)  A simplified schematic of Fig.~\ref{fig:circuit}, introducing symbols used in panels (b)-(d) and in Fig. \ref{fig:cartoon}.  The rectangle with a rectangular cut-out represents the 3d-cavity, the circle the qubit (in state $|\psi\rangle$), and the rectangle with a triangle cut-out the Kerr-resonator.  Object colors are coarse representations of the center frequency of each component.  (b) An empty 3d-cavity is driven by a resonant microwave drive (left-facing blue arrow), microwave power is built up in the 3d-cavity (cut-out is filled in blue), and the drive is reflected (right-facing arrow).  When the same cavity is driven by a slightly-off- (light blue arrows) or far-off-resonant drive (green arrows), less or no power is built up inside the cavity. The reflected signal will now have a phase that depends on the sign of the detuned incident field (indicated by an either dotted or dashed left-facing arrow). (c) When the total power incident on the Kerr-resonator is sufficiently low the power built up is relatively low (lightly-filled triangle) but increases disproportionally when a certain pair of critical powers, $|\alpha_{c\pm}|^2$, are exceeded as detailed in the text (fully-filled triangle).  (d)  The qubit inside the 3d-cavity will shift the center frequency of the cavity by a small, but significant amount, as described in the text and yields a qubit-state dependent phase shift of the reflected signal.  When the power is built up in the 3d-cavity, the center frequency of the qubit shifts proportionally (yellow circle accompanying the filled-in cut-out).} \label{fig:Mechcartoon}
\end{figure}

\begin{figure*}
\includegraphics[width=\linewidth]{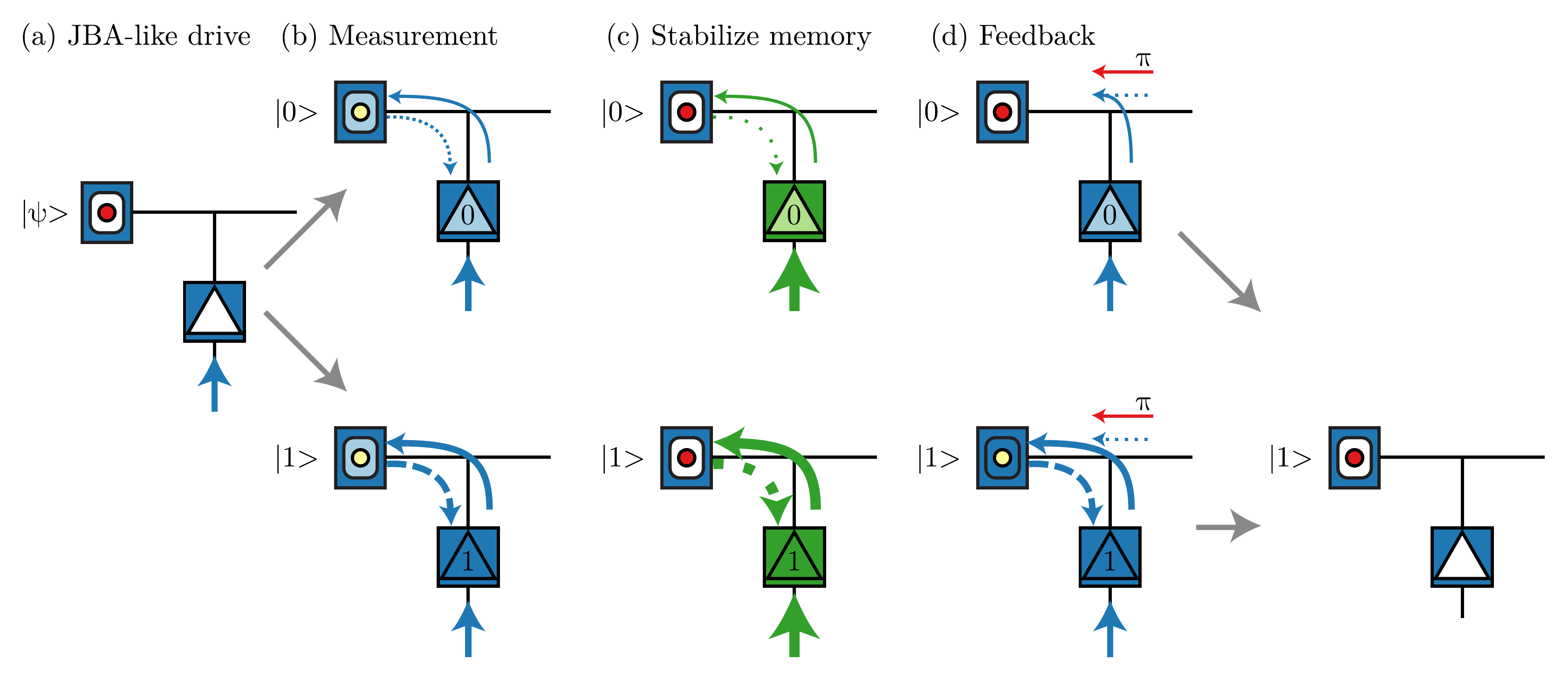}
\caption{(Color online) Protocol for measurement, classical memory storage, and feedback to a qubit.  a)  With the qubit in an initial state $|\psi\rangle$, the Kerr-resonator and drive are tuned to be co-resonant with the 3d-cavity (all three are colored blue).  (b) Measurement operation as described in the text.  (Note: the signal emitted by the Kerr-resonator is typically phase shifted as well, which we disregard in this section for simplicity.) (c) Stabilizing the classical memory state by tuning the Kerr-resonator and drive, and increasing drive power.  (d)  Using the classical memory state to apply a conditional $\pi$-pulse to the qubit, stabilizing it in the $|1\rangle$ state.} \label{fig:cartoon}
\end{figure*}

\begin{figure}[b]
\includegraphics[width=\linewidth]{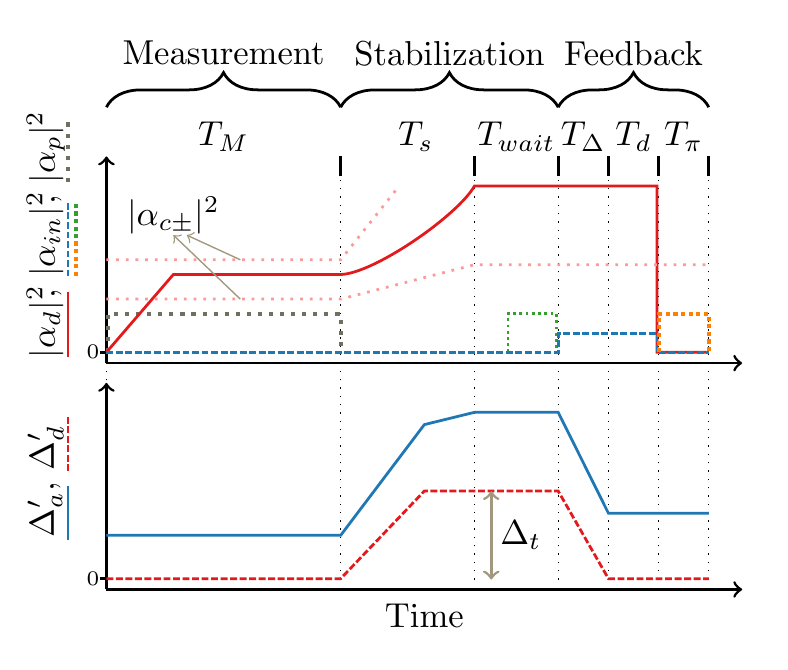}
\caption{(Color online) Illustration of the external controls used in the feedback scheme. The upper panel sketches the power of the input drives, while the lower panel illustrates the detuning between the Kerr-resonator frequency and the 3d-cavity, $\Delta_a' = \omega_a - \omega_b$ (solid blue line) and between the drive and the 3d-cavity $\Delta_d' = \omega_d - \omega_b$ (dashed red line). During the first time-step, $T_M$, we apply a measurement driven by $\alpha_d$ (solid red line) and an drive at the 3d-cavity, $\alpha_p$ (dashed gray line), which is followed by a stabilization time $T_s$, where the drive and Kerr-resonator parameters are changed to further stabilize the Kerr-resonator states by placing the drive far away from the critical input powers indicated by $\alpha_{c\pm}$ (thinly dotted light red line). Simultanously the Kerr-resonator and the input-drive are tuned away from the 3d-cavity by $\Delta_t$. After a waiting time $T_{wait}$ in which no parameters are changed, we tune the Kerr-resonator back into resonance with the 3d-cavity. During the second $T_\Delta$ a classical input drive, $\alpha_{in}$ (dashed blue line) is applied to the upper port and will interfere with the field from the Kerr-resonator. Finally, the $\pi$-pulse (dotted orange line) of duration $T_\pi$ excites the qubit state depending on the Kerr-resonator field strength. An optional pi-pulse (dotted green line) at the end of $T_{wait}$ restores the qubit to its measured state, as described in the final paragraph of Sec.~\ref{sec:Scheme}.} \label{fig:scheme}
\end{figure}

The basic operations that the network exploits are depicted schematically in Fig.~\ref{fig:Mechcartoon}.  When a resonant mode, like an empty 3d-cavity in Fig.~\ref{fig:Mechcartoon}~(b), is driven by a slightly off-resonant microwave signal, microwave power builds up in the mode and the signal is reflected with a certain phase shift.

While the center frequency of the 3d-cavity is fixed, the Kerr-resonator's center frequency is highly dynamic. Firstly, applying an external flux bias through the Kerr-resonator's dc superconducting quantum interference devices (SQUIDs) via the flux control line will shift the center frequency\cite{castellanos2008amplification}. Secondly, the amount of power circulating inside the Kerr-resonator also decreases the center frequency. This power-induced shift is a highly non-linear function of the externally applied, near-resonant microwave drives (see Sec.~\ref{sec:jba}), and Fig.~\ref{fig:Mechcartoon} (c) depicts the critical function for the purpose of this scheme. If the total incident power of red-detuned microwave drives is below a pair of detuning-dependent critical drive powers, $|\alpha_{c\pm}|^2$, the frequency shift will be negligible and little power is built up in the Kerr-resonator. If the incident power exceeds these critical values, though, the center frequency decreases significantly and it causes a stronger internal power to build up inside the resonator. If the incident power lies between the two critical values, the Kerr-resonator's center frequency and internal power are bistable and hysteric (again, see Sec.~\ref{sec:jba} for details).

In Fig.~\ref{fig:Mechcartoon} (d) we depict the inclusion of a qubit, which changes the behavior of the 3d-cavity. The qubit state shifts the center frequency of the 3d-cavity and, consequently, the microwave signal from the 3d-cavity has one phase shift when the qubit is in the $|0\rangle$ state and another if it is in the $|1\rangle$ state. Conversely, the center frequency of the qubit is shifted in proportion to the amount of power built up in the 3d cavity. The details will be considered in Sec.~\ref{sec:full}.

Figure~\ref{fig:cartoon} depicts a protocol that uses these functions to measure the qubit, store the result in a classical memory, and apply conditional feedback to stabilize the $|1\rangle$ state. In Fig.~\ref{fig:cartoon} (a), the qubit is initially in an arbitrary state $|\psi\rangle$, the Kerr-resonator is tuned using the flux control line to be near-resonant with the 3d-cavity, and the Kerr-resonator is driven through its bottom, weakly-coupled port. Some amount of drive power passes through the Kerr-resonator and drives the 3d-cavity, Fig.~\ref{fig:cartoon} (b). The signal reflected by the 3d-cavity is partially lost in the remaining tee-port (this loss is not depicted) and partially reflected back to the Kerr-resonator. The phase of this reflected signal depends on the qubit state and interferes with the Kerr-resonator drive.  Specifically, a $|1\rangle$ ($|0\rangle$) qubit state will reflect a signal that interferes constructively (destructively) with the Kerr-resonator drive and pushes the total incident power above (below) the Kerr-resonators critical values. A disproportionally high (low) internal power then builds up in the Kerr-resonator.  Thus, the system measures the qubit state and stores the result in the internal power level of the Kerr resonator.  A high or low power level represents a 1 or 0 classical ``memory'' state, respectively.

Figure \ref{fig:scheme} accompanies Fig.~\ref{fig:cartoon} and depicts the detunings and amplitudes of the external controls used in the protocol.  The measurement just described occurs over a time $T_M$. During this interval, the Kerr resonator/3d-cavity detuning $\Delta_a' \equiv \omega_a - \omega_b$ is small, while the drive/3d-cavity detuning $\Delta_d' \equiv \omega_d - \omega_b$ is zero with $\omega_a$,  $\omega_b$ and $\omega_d$ as the frequency of the Kerr-resonator, the 3d-cavity and the drive respectively. Furthermore, we depict in Fig. \ref{fig:scheme} a drive applied at the Purcell-protected port, which adds additional signal of the qubit state to the Kerr-resonator but does not change the qualitative  behaviour depicted in Fig. 3. Also shown in Fig. \ref{fig:scheme} is the drive field of the Kerr-resonator, $\alpha_{d}$, which is slowly increased to a value between the two critical values, $\alpha_{c}$, between which the Kerr-resonator is bistable (see sec.~\ref{sec:jba}).

This memory state can be protected against subsequent changes in the qubit state by increasing the drive power and detuning the drive frequency and Kerr-resonator center frequency in tandem, as indicated by the changed color in Fig.~ \ref{fig:cartoon} (c). Although the phase of the drive reflected by the 3d-cavity is now almost independent of the qubit state, the total power incident on the Kerr-resonator remains relatively high or low depending on the most recent memory result. The classical memory state has been effectively stored and may be retained indefinitely. The external controls applied during stabilization is depicted in Fig.~\ref{fig:scheme}. First, during a stabilization time $T_s$, the detuning of the Kerr-resonator and the drive amplitude are increased, which further separates the two critical values.  This ensures that neither fluctuations in the field nor subsequent qubit dynamics will induce changes in the classical memory state.  To further decouple the classical memory from the qubit, the Kerr-resonator and drive are detuned from the 3d-cavity by a large amount, $\Delta_t$ simultaneously with the stabilization.  At this point, the memory may be held for an extended time $T_{wait}$.

Finally, Fig.~\ref{fig:cartoon} (d) depicts how the qubit may be deterministically prepared in the $|1\rangle$ state, regardless of the initial state $|\psi\rangle$. The Kerr-resonator and drive are tuned back toward the 3d-cavity's frequency, resuming their values from the measurement step (Fig.~\ref{fig:cartoon} (b) ).  This is depicted in Fig.~\ref{fig:scheme} in which the detunings are brought back to their  original values.  If the Kerr-resonator memory state is 0, the power incident on the 3d-cavity through the Kerr-resonator is weak and can be eliminated by superimposing it with another weak and phase shifted drive through the remaining tee-port, depicted in Fig.~\ref{fig:cartoon} (d) and by the dashed blue line in Fig.~\ref{fig:scheme}. In this case, no power is built up in the 3d-cavity and the qubit center frequency remains un-shifted. If the Kerr-resonator's memory state is 1, however, the drive incident on the 3d-cavity is relatively large and some power will build up in the cavity  (despite the phase-shifted drive through the remaining tee-port) and shift the qubit center frequency. We are therefore left with a qubit whose center frequency is either shifted or not, depending on the classical memory state. If a third microwave pulse tuned to apply a $\pi$-rotation at the original qubit frequency is now applied at the tee-port (Fig.~\ref{fig:cartoon} (d) and the dotted orange line of duration $T_\pi$ in Fig.~\ref{fig:scheme}), a $|0\rangle$ qubit will be rotated to $|1\rangle$, but a $|1\rangle$ qubit will be off-resonant and will remain in $|1\rangle$.  Thus, a qubit in the $|1\rangle$ state is prepared deterministically.

The protocol closes an autonomous feedback loop at cryogenic temperatures but from the point of view of an operator at room temperature, it constitutes an open loop protocol.  This is conceptually similar to quantum reservoir engineering schemes \cite{geerlings2013demonstrating, PhysRevLett.109.183602, shankar2013autonomously, holland2015single}, in the way that it changes the effective ground state and does not rely upon any conditional room temperature operations common to conventional feedback schemes \cite{campagne2013persistent, riste2012feedback, riste2013deterministic, vijay2012stabilizing}. The protocol, however, relies upon the performance of a coupling to a dissipative meter system and, thus, it falls conceptually in-between conventional feedback and quantum reservoir engineering schemes.

We can further demonstrate that the classical memory may be usefully retained for an arbitrary amount of time.  For example, the setup proposed here can return the classical state stored in the memory to the qubit, even after the qubit has undergone $T_1$ decay to $|0\rangle$ (and assuming no spontaneous excitation to $|1\rangle$ for demonstration purposes), i.e. $T_{wait}\gg T_1$.  Just before we bring the Kerr-resonator back onto resonance, we apply a $\pi$-pulse on the bare qubit frequency bringing the qubit into the excited state, dotted green line in Fig.~\ref{fig:scheme}. After this pulse, we apply the same input field and tuning sequence as above. A resonant $\pi$-pulse will now prepare the qubit in the excited state if and only if that was the state measured earlier.

\section{Quantitative analysis of the full system}
\label{sec:full}

The circuit illustrated in Fig. \ref{fig:circuit} consists of different components. We shall now provide the formalism for the quantitative description of the systems and their dynamics. We will begin with a breif review of the readout mechanism followed by a complete analysis of the full system.

\subsection{The Josephson bifurcation readout}
\label{sec:jba}

Since the readout is fundamental for the operation of the scheme we will briefly review how a Kerr-resonator performs a projective readout of a qubit. A Kerr-resonator may be implemented as an array of SQUIDS in parallel with a capacitance \cite{castellanos2008amplification}. The SQUIDS provide the inductance for the equivalent $LC$-circuit and they permit tuning of the resonance of the Kerr-resonator by a magnetic flux. While a single SQUID provides a Kerr-nonlinear behaviour, an array will dilute the non-linearity and allow a much higher photon number in the Kerr-resonator \cite{eichler2014controlling}. The resonator loss rates associated with the two coupling capacitors in Fig. \ref{fig:circuit} are $\kappa_d$ and $\kappa_a$ and we assume a classical input field $\alpha_d$ coupled with strength $\sqrt\kappa_d$ to the resonator mode. In a frame rotating with the drive frequency, $\omega_d$, the classical field amplitude in the resonator solves the non-linear equation,
\begin{equation}
\dot{\alpha}=-i\Delta_a \alpha -iK|\alpha|^2\alpha - \frac{\kappa_a+\kappa_d}{2}\alpha-\sqrt{\kappa_d}\alpha_d, \label{eq:simple}
\end{equation}
where $K$ denotes the Kerr-nonlinearity, which originates from the non-linear inductance of the SQUID-array.

In the classical field approximation, the driven Kerr-resonator relaxes into a steady-state which is found by setting $\dot{\alpha} = 0$ in Eq. \eqref{eq:simple}. This yields an equation for $\alpha_d$ in terms of $\alpha$. We are, at present, not concerned with the phase of the solution, and by multiplying the expression for $\alpha_d$ with its complex conjugate, we find the non-linear relation between the driving field and the intracavity field,
\begin{align}
\kappa_d |\alpha_d|^2 &= \Delta_a^2 |\alpha|^2 + 2\Delta_a K |\alpha|^4  \nonumber\\&\quad+ K^2 |\alpha|^6+ \frac{(\kappa_a + \kappa_d)^2}{4}|\alpha|^2. \label{eq:a_ss}
\end{align}
This expression yields unique values of $|\alpha_d|^2$ for any $\alpha$, while multiple values of the intra-cavity field strength $\alpha$ may be compatible with the same driving field. We analyze this possibility by solving $\frac{d |\alpha_d|^2}{d |\alpha|^2}=0$, which yields the critical photon numbers $n_{c\pm}=|\alpha_{c\pm}|^2$,
\begin{align}
 n_{c\pm} = - \frac{2\Delta_a K}{3 K^2} \Bigg( 1 \mp \sqrt{1 - \frac{3(\Delta_a^2 + (\kappa_a + \kappa_d)^2/4)}{4\Delta_a^2}} \Bigg). \label{eq:nc}
\end{align}

\begin{figure}[t]
\includegraphics[width=\linewidth]{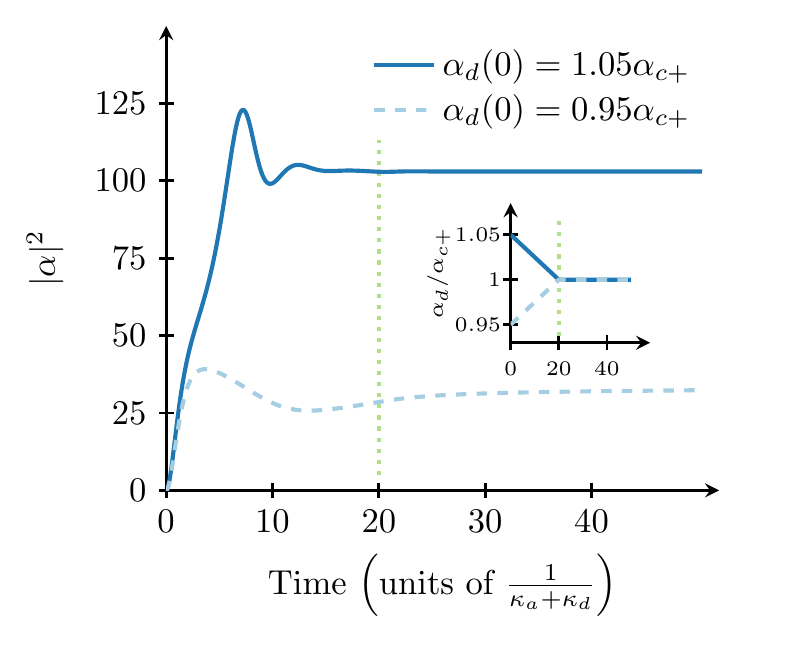}
\caption{(Color online) Solution of the classical field equation \eqref{eq:simple} with drive amplitudes slightly above and slightly below the critical drive, $\alpha_{c\pm}$. The inset shows how the two applied drive amplitudes linearly approach the critical value at time $20/(\kappa_a + \kappa_d)$ indicated by the vertical dotted (green) line. A constant detuning $\Delta_a = 1.75\Delta_{a,c}$ is assumed between the drive and the Kerr-resonator and the non-linearity is set to $K=-0.012\Delta_a$. For the slightly higher drive power, we obtain a steady state with ${\sim}110$ photons, which is much larger than the ${\sim}35$ photons obtained for the only slightly weaker inital drive. } \label{fig:jba}
\end{figure}

If $\Delta_a = \Delta_{a,c} \equiv \sqrt{3/4} (\kappa_a + \kappa_d)$, we obtain a point of inflection in the equation (\ref{eq:a_ss}), while if we increase $\Delta_a$ beyond $\Delta_{a,c}$ we obtain two unique solutions, $n_{c\pm}$. For driving strengths between the two corresponding field values, Eq. \eqref{eq:a_ss} has one unstable and two stable solutions. The Kerr-resonator bifurcated in the same way as a JBA, which derives its name from this bifurcation, and now a very weak initial perturbation is sufficient to make the resonator field attain different stationary solutions that depend on this perturbation. The Kerr-resonator will latch into one of these two solutions even when the perturbation is switched off. In Fig. \ref{fig:jba} we see an example of how this bifurcation manifests itself as a very large difference in photon number with only a small difference in input power. This large difference persists even when the input power is levelled out. The perturbation could be introduced in the upper input port to the Kerr-resonator, for example, from the 3d-cavity controlled by the qubit state in Fig.~\ref{fig:circuit}. In this case, a high fidelity QND detection of the qubit state results if the output fields are amplified and transferred to room temperature electronics \cite{mallet2009single, vijay2009invited}. In this work, however, rather than transferring the output to room temperature, we will use the fact that the steady states of the Kerr-resonator are already classical states with high fidelity correlation with the qubit state.

The time scale for readout is naturally related to the time of bifurcation, which in general is complicated to predict but is associated with $1/K$. If the Kerr-resonator can bifurcate on a very short time scale, the readout fidelity might be poor. This is because the strong coupling between the Kerr-resonator and the 3d-cavity, needed for fast readout, will lead to a larger Purcell-decay of the qubit and a very high photon number inside the 3d-cavity, which will cause the measurement to be non-QND. This  obstacle can be mitigated by the Purcell-filtered port of Fig. \ref{fig:circuit}, allowing the cavity field to decay but maintaining the qubit lifetime. We can, thus, achieve very strong coupling and still remain in the QND-regime by effectively increasing the cavity decay rate. Eventually, though, the readout-speed and fidelity will still be limited by the dispersive coupling between the 3d-cavity and the qubit.

\subsection{Circuit analysis}

The Kerr-resonator and the qubit in the 3d-cavity are described by the Hamiltonians \cite{eichler2014controlling, PhysRevLett.108.240502}:
\begin{align}
H_K &= \hbar \omega_a a\dag a + \hbar \frac{K}{2} a\dag a a\dag a,\\
H_{cav} &= \hbar \omega_b b\dag b + \hbar \frac{\Omega_{qb}}{2} \sigma_z + \hbar\chi \, b\dag b \, \sigma_z, \label{eq:Hqbcav}
\end{align}
with resonator eigenfrequencies $\omega_{a,b}$, a qubit frequency $\Omega_{qb}$ and a dispersive interaction between the qubit and the 3d-cavity field with the strength $\chi$. Here $\sigma_z$ denotes the qubit operator such that $\sigma_z \ket{1} = \ket{1}$ and $\sigma_z \ket{0} = - \ket{0}$. The operators $a$ ($a\dag$) and $b$ ($b\dag$) describes the annihilation (creation) operators for the Kerr-resonator and the 3d-cavity respectively.

The Kerr-resonator and the qubit-cavity are connected by one-dimensional waveguides. The position dependent phase operator, $\phi(x,t)$, (time integral of the voltage) and charge density, $q(x,t)$, along the waveguides obey wave-equations and they can be expanded on left and right propagating wave solutions \cite{peropadre2013scattering, lalumiere2013input},
\begin{align}
\overlr{\phi}(x,t) &= \sqrt{\frac{\hbar Z_0}{4 \pi}} \int_0^{\infty} \frac{d\omega}{\sqrt{\omega}} \, \overlr{a}_\omega e^{-i(\omega t \pm k_\omega x)} + h.c. \\
\overlr{q}(x,t) &= -i\sqrt{\frac{\hbar Z_0}{4 \pi}} \int_0^{\infty} \hspace{-0.1cm} d\omega \, \sqrt{\omega}\, \overlr{a}_\omega e^{-i(\omega t \pm k_\omega x)} + h.c.
\end{align}
such that
\begin{align}
\phi(x,t) = \overleftarrow{\phi}(x,t) + \overrightarrow{\phi}(x,t).
\end{align}
In the above equations we have assumed that the wave-guide impedance $Z_0$ is constant and the same for all lines. We choose the origin of the position variable to be at the tee-junction. The left (right) arrow denotes the field moving towards (away from) the 3d-cavity and down (up) towards (away from) the Kerr-resonator.

The waveguides are joined by a tee-junction, and Kirchoff's equations yield \cite{pozar2009microwave}
\begin{align}
\phi_{out}(t) = \frac{1}{3}\phi_{in}(t) + \frac{2}{3}\overrightarrow{\phi}_K(0,t) + \frac{2}{3}\overrightarrow{\phi}_{cav}(0,t) \label{eq:bc-T}
\end{align}
for the reflected component, $\phi_{out}$, in terms of the waves incident on the junction, $\phi_{in}$. Now $\phi_{out}(t)$ depends, in particular, on the fields from the Kerr-resonator, $\overrightarrow{\phi}_K$, and from the 3d-cavity, $\overrightarrow{\phi}_{cav}$.

Let $q_{b}$ denote the charge degree of freedom at the 3d-cavity and similarly let $q_a$ denote the charge at the Kerr-resonator, such that we obtain the boundary conditions \cite{peropadre2013scattering}
\begin{align}
\overrightarrow{\phi}_K(d_K,t) &=\overleftarrow{\phi}_K(d_K,t) + Z_0 \beta_a q_{a}(t), \label{eq:bc-a} \\
\overrightarrow{\phi}_{cav}(d_{cav},t) &=\overleftarrow{\phi}_{cav}(d_{cav},t) + Z_0 \beta_b q_{b}(t), \label{eq:bc-b}
\end{align}
where $d_K$ and $d_{cav}$ denote the lengths of the waveguides from the junction to the Kerr-resonator and the 3D cavity respectively. The quantities $\beta_a$ and $\beta_b$ are unit-less geometric constants that depend on the relation between the effective coupling capacitance and the total capacitance of each subsystem. The charge variables $q_a$ and $q_b$, are given by the cavity mode operators $(a, a\dag)$ and $(b, b\dag)$, respectively. Through the so-called black-box quantization \cite{PhysRevLett.108.240502} we assume the Hamiltonian of the qubit-cavity system to be of the form Eq. \eqref{eq:Hqbcav}. Following this description, $q_b$ now also includes a contribution from the qubit and this exactly leads to the well-known Purcell-decay of the qubit. We will omit the qubit dynamics for now, but it will be considered in Sec. \ref{sec:num}.

An input-output relation similar to Eq. \eqref{eq:bc-b} can also be written for the lower output port of the 3d-cavity shown in Fig \ref{fig:circuit}. However, unlike the upper port, we will attach a Purcell filter \cite{PhysRevLett.112.190504, reed2010purcell, bronn2015reducing, sete2015quantum} to the output port. The sole purpose of this additional decay-channel for the 3d-cavity is to allow a very strong interaction between the Kerr-resonator and the 3d-cavity without exciting the 3d-cavity beyond the dispersive regime \cite{boissonneault2009dispersive}. The extra cavity decay does therefore not by itself improve the qubit readout, but it allows for a much stronger coupling between the Kerr-resonator and the 3d-cavity. So as long as the total decay rate of the 3d-cavity remains in the same order of magnitude as the dispersive coupling to the qubit, we can increase the coupling to the Kerr-resonator by increasing the decay through the Purcell-fitered channel by the same amount. In its simplest form the Purcell filter is made by a transmission line stub connected to the output line. This stub acts as a $\lambda/4$ impedance transducer, but only at a single frequency, $\omega_f$. Now, if $\omega_f = \Omega_{qb}$, the filter will short out the environment of the transmission line and thus create an effectively vanishing density of states in the environment at the qubit frequency, leading to vanishing Purcell-decay of the qubit through the filtered channel. The 3d-cavity, however, operates at a different frequency and is allowed to decay through the filtered channel.

We will assume a narrow frequency band of the travelling fields, $k_\omega \approx k$, and we restrict the analysis to the situation of wave propagation times much shorter than other dynamical time scales of the systems. This allows elimination of the wave-guide observables, $\phi(x,t)$ and $q(x,t)$, and we obtain the Heisenberg-Langevin equations for the $a$ and $b$ operators
\begin{align}
\dot{a} &= -i \omega_a a - i K a^{\dagger} a a - \frac{\kappa_a + \kappa_d}{2} a \nonumber\\ &\quad+ \sqrt{\kappa_a \kappa_b} b e^{i ( \phi_b - \phi_a )} - \sqrt{\kappa_a} a_{in} e^{-i\phi_a} - \sqrt{\kappa_d} a_d, \label{eq:hl-a}\\
\dot{b} &= -i \omega_b b -i \chi \sigma_z b - \frac{\kappa_b + \kappa_p}{2} b \nonumber\\ &\quad+ \sqrt{\kappa_a \kappa_b} a e^{i (\phi_a - \phi_b)}  - \sqrt{\kappa_b} a_{in} e^{-i\phi_b} - \sqrt{\kappa_p} a_p.\label{eq:hl-b}
\end{align}
where the 3d-cavity and the Kerr resonator are coupled with the strength $\sqrt{\kappa_a \kappa_b}$. In Eqs. \eqref{eq:hl-a} and \eqref{eq:hl-b}, we have also included an extra input and decay channel for the Kerr resonator corresponding to the lower input port of Fig.1 and for the 3d-cavity through the upper port ($\kappa_b$) and the Purcell filtered lower port ($\kappa_p$). The operators $a_{in}$ and $a_d$ are the input field operators, while the phases $\phi_a = d_K k$ and $\phi_b = d_{cav} k$ come from the propagation through the transmission lines to the Kerr-resonator and the 3d-cavity.

Having Eqs. \eqref{eq:hl-a} and \eqref{eq:hl-b} we can in principle solve the full dynamics of the system, but since $a$ and $b$ are operators and the mean value of $a$ might be on the order of ${\sim}10$, we would need to expand the Hilbert space on a large number of Fock states to solve the full quantum dynamics. We, therefore, employ a stochastic mean-field description by making the c-number substitutions:
\begin{align}
&a \rightarrow \alpha e^{-i\omega_d t}, && && b \rightarrow \beta e^{-i\omega_d t}, \nonumber\\
&a_{in} \rightarrow \alpha_{in}e^{-i\omega_d t}, && && a_d \rightarrow \alpha_d e^{-i\omega_d t}, \nonumber\\
&a_p \rightarrow \alpha_p \,e^{-i\omega_d t}, && && \sigma_z \rightarrow \langle \sigma_z\rangle. \nonumber
\end{align}
For convenience we have transformed our equations to a frame rotating at the frequency of the external drive, $\omega_d$, which will make the detunings $\Delta_a = \omega_a - \omega_d$ and $\Delta_b = \omega_b - \omega_d$ appear in the equations of motion. We can make the complex field variables represent random samplings of the quantum Wigner quasiprobability distributions if we include stochastic terms, representing the quantum noise associated with the dissipation terms in  \eqref{eq:hl-a} and \eqref{eq:hl-b}, \cite{gardiner2004quantum}. 

Following the prescription in \cite{gardiner2004quantum}, we obtain fluctuating complex noise contributions 
\begin{align}
\xi_a (t) &= \sqrt{\frac{1}{2}(\kappa_a + \kappa_d) (\bar{N} + \frac{1}{2})} dW_{a,1} \nonumber\\ &\quad+ i \sqrt{\frac{1}{2}(\kappa_a + \kappa_d) (\bar{N} + \frac{1}{2})} dW_{a,2},\label{eq:xia} \\
\xi_b (t) &= \sqrt{\frac{1}{2}(\kappa_b + \kappa_p) (\bar{N} + \frac{1}{2})} dW_{b,1} \nonumber\\ &\quad+ i \sqrt{\frac{1}{2}(\kappa_b + \kappa_p) (\bar{N} + \frac{1}{2})} dW_{b,2},\label{eq:xib}
\end{align}
where $dW_{i,n}$ denote independent Wiener processes with $\exv{dW_{i,n}} = 0$ and $\exv{dW_{i,n}^2} = dt$ and $\bar{N}$ is the mean photon number in the thermal bath coupled to the resonators.

Adding these terms to the deterministic mean value equations, we obtain the equations of motion for $\alpha$ and $\beta$,
\begin{align}
\dot{\alpha} &= -i \Delta_a \alpha - i K |\alpha|^2\alpha - \frac{\kappa_a + \kappa_d}{2} \alpha + \sqrt{\kappa_a \kappa_b} \beta e^{i ( \theta_b - \theta_a )} \nonumber\\ &\quad- \sqrt{\kappa_a} \alpha_{in} e^{-i\theta_a} - \sqrt{\kappa_d} \alpha_d + \xi_a(t), \label{eq:alpha}\\
\dot{\beta} &= -i \Delta_b \beta -i \chi \exv{\sigma_z} \beta - \frac{\kappa_b+\kappa_p}{2} \beta + \sqrt{\kappa_a \kappa_b} \alpha e^{i (\theta_a - \theta_b)}  \nonumber\\ &\quad- \sqrt{\kappa_b} \alpha_{in} e^{-i\theta_b} - \sqrt{\kappa_p} \alpha_p + \xi_b(t), \label{eq:beta}
\end{align}
The replacement of the full quantum operator equations by noisy mean field equations has been succesfully applied in a wide range of quantum optics problems, and due to the weak Kerr non-linearity we expect this description to correctly describe the mean value of the resonator field as long as the qubit is in one of the eigenstates of $\sigma_z$. We refer to \cite{PhysRevApplied.1.054005} for a recent comparison of full quantum and noisy mean field calculations in an optical network system with Kerr-nonlinearities. This system also shows bistability and switching dynamics, similar to ours and shows a fully satisfactory agreement between the methods. In the appendix we verify this agreement for simple instances of our problems, with the only discrepancies concerning the quantitative dynamics just at the bifurcation threshold, a minor concern, since the precise properties have to be calibrated in the experiments.

As one sees from Eq. \eqref{eq:beta}, the qubit state influences the driving of the 3d-cavity, which in turn affects the driving of the Kerr-resonator. In the regime where the qubit-dependent feedback drives the Kerr-resonator into two different stable states, this will lead to a projective measurement of the qubit, and storage of the outcome in the Kerr-resonator. The tunability of both $\Delta_a$, $\alpha_d$ and $\alpha_{in}$ now enables the implementation of the schemes described in Sec. \ref{sec:Scheme}.

\section{Numerical simulations}
\label{sec:num}

So far we have treated the qubit as a semi-classical object with two states, $\ket{0}$ and $\ket{1}$, each of which shifts the frequency of the 3d-cavity. In reality the qubit is described by the quantum mechanical density matrix, $\rho_{qb} = \sum_{i,j = 0,1} \rho_{i,j} \ket{i}\bra{j}$, and its dynamics are described by a master equation.

However, we choose to describe the Kerr-resonator as well as the 3d-cavity by stochastic complex fields, where a single trajectory is determined by the particular instance of $dW(t)$ restricting ourselves to classical-like states. Similarly we can unravel the master equation for the qubit into a non-Hermitian Schrödinger equation \cite{dalibard1992wave, wiseman2009quantum}
\begin{align}
\frac{d}{dt} \vert \psi \rangle &= - i \frac{\Delta_{qb} + 2\chi |\beta|^2}{2} \sigma_z \vert \psi \rangle - \frac{\gamma}{2} \vert 1 \rangle \langle 1 \vert \,\vert \psi \rangle \nonumber \\ &\quad+ (\Omega_d(t) \sigma_+   + \Omega_d^*(t) \sigma_-  )\vert \psi \rangle, \label{eq:qubit}
\end{align}
where $\Delta_{qb} = \Omega_{qb} - \omega_d$ and $\Omega_d(t)$ describe a drive at the qubit frequency. The non-Hermitian Schrödinger equation is designed such that the norm of the wavefunction, $\ket{\psi}$ decays, due to the excited state decay rate $\gamma = 1/T_1$. When the norm becomes smaller than a random number $R$, drawn uniformly between 0 and 1, we apply the jump operator $\sigma_-$ to the state and then we renormalize $\ket{\psi}$. This procedure reproduces the average density matrix evolution of the qubit.

A superposition state of the qubit generally yields expectation values $\exv{\sigma_z} = \bra{\psi} \sigma_z \ket{\psi}$ that differ from the eigenvalues $\pm 1$, but after the bifurcation readout of the qubit, we expect the state of the coupled system to be in a completely mixed state with different field states pertaining to the excited and ground state qubit, $\rho = P_{0} \ket{0}\bra{0} \otimes \rho_{a,0} + P_1 \ket{1}\bra{1} \otimes \rho_{a,1}$, with $\rho_{a,0\,(1)}$ being the 0 (1) memory state of the Kerr-resonator. Here $P_n$ denote the probabilities for the qubit to be in state 0 and 1, and due to the QND interaction, these probabilities do not change during the readout. Thus, we can simulate the two classical qubit cases separately, and on average generate the full master equation evolution.

Now, we can investigate the performance of the proposed scheme by numerically solving Eqs. \eqref{eq:alpha}, \eqref{eq:beta} and \eqref{eq:qubit}, but with the qubit in an eigenstate as described above leaving $\exv{\sigma_z} = \pm 1$. For the numerical simulation we need realistic parameters, so we choose a detuning between the qubit and the resonator at $\Delta_{qb} = 2\pi \times 1.2$ GHz. Furthermore we assume that the coupling between the two systems are at $g = 2\pi \times 122$ MHz, which yields a dispersive shift at \mbox{$\chi = g^2 E_c / (\hbar \Delta_{qb} ( \Delta_{qb} - E_c / \hbar) )= 2\pi\times -2.5$ MHz} \cite{PhysRevA.76.042319} with the anharmonicity of the qubit given by \mbox{$E_c/\hbar = 2\pi \times 300$ MHz}. We set the decay-rate for the 3d-cavity to $\kappa_b = 2\pi \times 1$ MHz and $\kappa_p = 2\pi \times 4$ MHz. For the Kerr-resonator we assume $K = 2\pi \times (-0.4)$ MHz, $\kappa_a = 2\pi \times 5$ MHz and $\kappa_d = 2\pi \times 0.3$ MHz with a detuning of the center frequency from the drive initially at $\Delta_a = 3.5\times(\kappa_a + \kappa_d)$. Finally, for the qubit we assume the intrinsic decay-rate $\gamma_{qb} = 2\pi \times 5$ kHz and from the cavity we get an additional Purcell-rate of $\gamma_p = \kappa_b \frac{g^2}{\Delta_{qb}^2}$, which yields $\gamma = \gamma_{qb} + \gamma_p = 2\pi\times 15.3$ kHz corresponding to a qubit lifetime of $T_1 = 10$ $\mu$s.

For the bifurcation readout our goal is a large difference in the Kerr-resonator photon number conditioned upon the qubit state. We find that a measurement time of $T_M = 400$ ns is sufficient to obtain a significant difference between the signals for the different qubit states compared to the noise of the input field. A shorter measurement time would be dominated by noise and yield a poor readout fidelity. During the first 80 ns, we slowly increase the drive towards $\alpha_{d} = 33.02 \sqrt{\kappa_d}$, which is just below the larger $\alpha_{c\pm}$ set by Eq. \eqref{eq:nc}. The Purcell protected port was driven by $\sqrt{\kappa_p} \alpha_p/(2\pi i) = 8$ MHz, with the $i$ indicating a $\pi/2$-phase shift of the drive compared to the rest of the drives. The next step is the stabilization where we linearly increase $\Delta_a$ by a factor of 1.7 and $\alpha_d$ by a factor of 1.8. This increases the difference in amplitude between the two solutions of Eq. \eqref{eq:nc} and puts $\alpha_d$ further away from both -- as a consequence the probability of a spontaneous jump between the two solutions vanishes. To ensure that we remain on the right branch of the bifurcation curve $T_s$ cannot be too small and we choose $T_s = 150$ ns. The stabilization time $T_s$ might be chosen smaller if we apply a non-linear tuning scheme for $\Delta_a$ and $\alpha_d$, but with the linear increase we numerically find that smaller times introduce significant errors.

The next step is the frequency tuning of the Kerr-resonator along with the drive. To ensure that we are indeed very far-detuned from the 3d-cavity after the detuning, we choose $\Delta_t = 2\pi \times 30$ MHz and to avoid exciting the Kerr-resonator while tuning the frequency, we cannot exceed $d\Delta_a / dt = \kappa / \frac{1}{\kappa}$, with $\kappa = \kappa_a + \kappa_d$. This limits the detuning time to $T_\Delta = 100$ ns. This detuning between the Kerr-resonator and the 3d-cavity is also occurring during $T_s$ simultaneous with the increase in $\Delta_a$ (see Fig. \ref{fig:scheme}). The waiting time $T_{wait}$ can be chosen according to our applications. We then wait for a driving time $T_d = 15$ ns before the $\pi$-pulse with Rabi-frequency $\Omega_d = 2\pi \times 7$ MHz and \mbox{$T_\pi = \pi/(2\Omega_d) = 35.7$ ns}.

\begin{figure}[t]
\includegraphics[width=\linewidth]{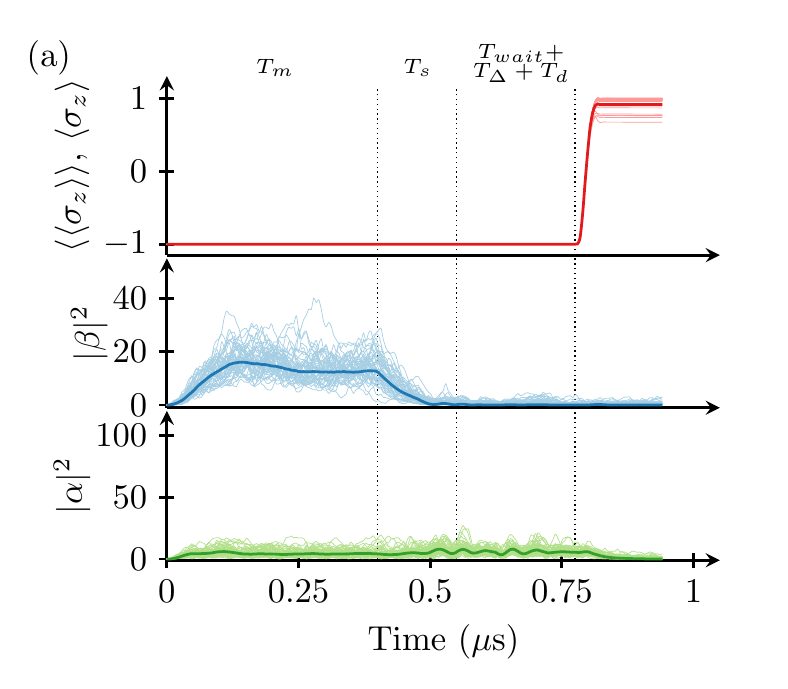} \vspace*{-1.2cm}

\includegraphics[width=\linewidth]{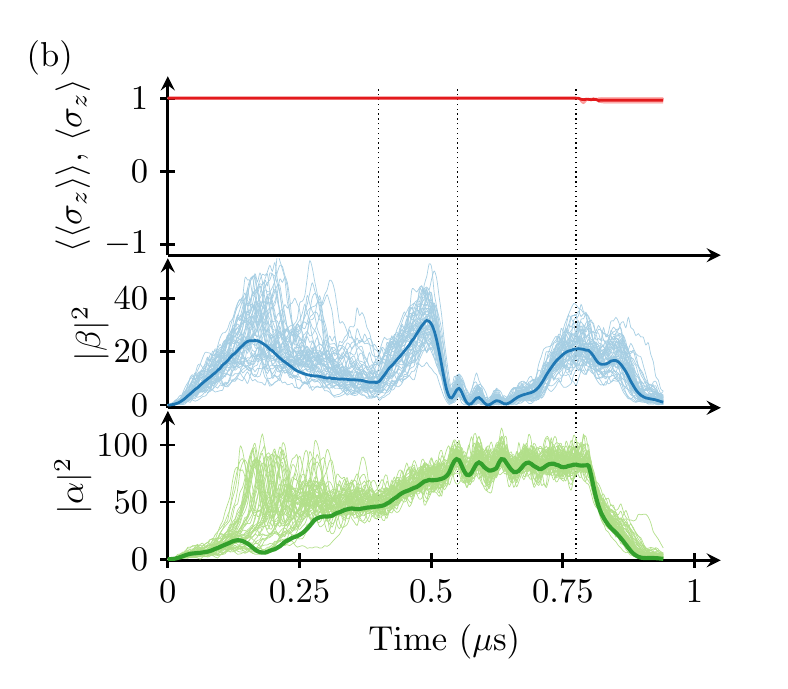}\vspace*{-0.8cm}
\caption{(Color online) Numerical simulations of the state preparation with parameters described in the text without qubit decay. In (a) we start with the qubit in $\ket{0}$ and bring the state into $\ket{1}$ with 95.7\% fidelity while in (b) we start with the qubit in the $\ket{1}$-state, where it remains with 98.5\% fidelity. The upper panel of each figure displays the mean value of $\sigma_z$, the middle panel displays the photon population in the 3d-cavity and the lower panel the population in the Kerr-resonator. In each panel we plot both the mean values for 50 trajectories (thin lines) and the average over 200 trajectories (thick line). The vertical dotted lines indicate the different time intervals similar to Fig. \ref{fig:scheme} and explained in the text.} \label{fig:stateprep}
\end{figure}

\begin{figure}[t]
\begin{flushleft}
\includegraphics[width=\linewidth]{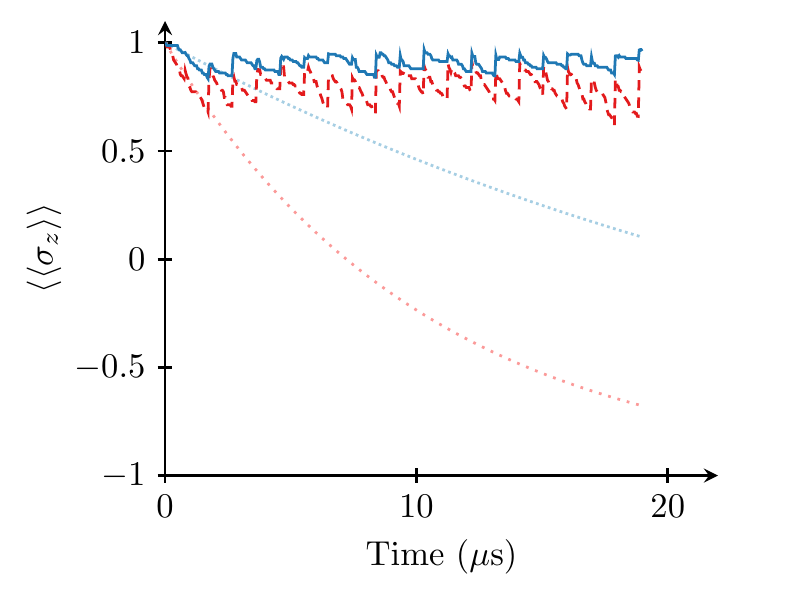}
\end{flushleft}
\vspace{-1cm}
\caption{(Color online) Stabilization of the qubit by running the same scheme as in Fig. \ref{fig:stateprep} but 20 times after each other. The parameters are all listed in the text. The dashed (red) line is the average of 400 trajectories, while the dotted (light red) line is an exponential Purcell-enhanced decay with the decay-rate of the qubit, $1/\gamma = 10$ $\mu$s. The solid (blue) curve shows the same but with a decay-rate of $1/\gamma = 30$ $\mu$s acheived with an additional Purcell-filter as explained in the text and the densely dotted (light blue) curve shows the corresponding exponential decay.} \label{fig:stabilization}

\end{figure}

\subsection{State preparation and stabilization}
For the state preparation scheme we want to have the waiting time, $T_{wait}$, as small as possible in order to quickly initialize the state into the excited state. Since $1/(\kappa_b+\kappa_p) = 31$ ns, the 3d-cavity is (almost) empty after the two tuning steps, so we can choose a very small waiting time $T_{wait} = 25$ ns. After $T_{wait}$ we tune the resonator and drive frequency back again and we also apply the input field $\alpha_{in}$, which is of equal amplitude but opposite phase of the 0 state of the Kerr-resonator obtained when the qubit was measured to be in its ground state. When we are back on resonance such that $\omega_d = \omega_b$, we apply the $\pi$-pulse and we obtain numerically a pulse fidelity of 98\%. The imperfection arises from the field fluctuations in the 3d-cavity and due to the fluctuating field emitted from the Kerr-resonator. These fluctuations are the same fluctuations that limit the readout fidelity. We apply square $\pi$-pulses in our simulations, and a study of pulse-shaping is beyond the scope of the present work, but we note that more elaborate pulse shapes may prevent leakage of population into higher excited states of the weakly anharmonic transmon qubits \cite{PhysRevLett.103.110501} and to mitigate some types of quantum \mbox{fluctuations \cite{PhysRevLett.99.170501}}.

In Fig. \ref{fig:stateprep} we show the evolution of $|\beta|^2$  and from the numerical simulation we achieve $|\beta|^2 \approx 0.5 n_{crit}$ during the readout, where $n_{crit} = \Delta^2/(4g^2)$ is the critical photon number at which the dispersive regime breaks down \cite{boissonneault2009dispersive}. The effective detuning of the qubit due to higher order effects is reduced by around 10\% at $0.5 n_{crit}$ (included in numerical calculations). We cannot improve the scheme by increasing the drive power, as we are already close to the maximum photon number where the dispersive coupling offers a valid description. We even notice that just after the bifurcation, the photon number is momentarily increased beyond the dispersive regime. While this is not important for the readout since we have already transferred the state to the resonator, we need to take into account the so-called dressed dephasing \cite{PhysRevA.77.060305}, which adds an additional decoherence channel for the qubit proportional to the photon number in the 3d-cavity. This effect will effectively decrease the lifetime of the qubit state and since the typical time where each trajectory remains above $n_{crit}$ is 25 ns, we expect a 1\%-2\% additional error from the additional dephasing. In \cite{PhysRevA.77.060305} it is also shown that above a saturation point around 0.5$n_{crit}$, dispersive readout cannot be improved, however higher power leads to a faster bifurcation and more stability of the bifurcated states, so the optimal power of our scheme is expected to be much higher than just the optimal power for pure dispersive readout. If there is still a demand for lower photon number, around 0.1-0.2$n_{crit}$, our scheme will still work, but with a fidelity of only around 85\%-90\%.

Figure \ref{fig:stateprep} demonstrates the ability of our scheme to deterministically prepare the qubit in the exited state from any unknown state. For simplicity we have not included qubit decay in these simulations. Starting in the excited state yields a succes probability of 98.1\% and if the qubit starts in the ground start we achieve a success probability of 96.1\%. The error originates partly from the imperfect $\pi$-pulse and partly from the fact that the readout-fidelity is only around 98\%.

Qubit decay is included with a finite qubit lifetimes of 10 $\mu$s (red line) and 30 $\mu$s (blue line)  in the simulations shown for 20 runs in Fig. \ref{fig:stabilization} with the same parameters as in Fig. \ref{fig:stateprep}. The time averaged state fidelity is 89.9\% and it can be increased to 93.5\% by averaging only at the instants of the protocol, where the preparation scheme is completed and the fidelity is known to be highest. The main limitation for the stabilization is the assumed qubit lifetime.  For example, installing an additional Purcell-filter on the upper output port (possibly between the 3d-cavity and the tee-junction) increases the qubit lifetime to around 30 $\mu$s and leads to an average fidelity with the excited state of 95.4\% (and 96.8\% by only averaging after each protocol ends) in our calculations.

\begin{figure}[t]
\includegraphics[width=\linewidth]{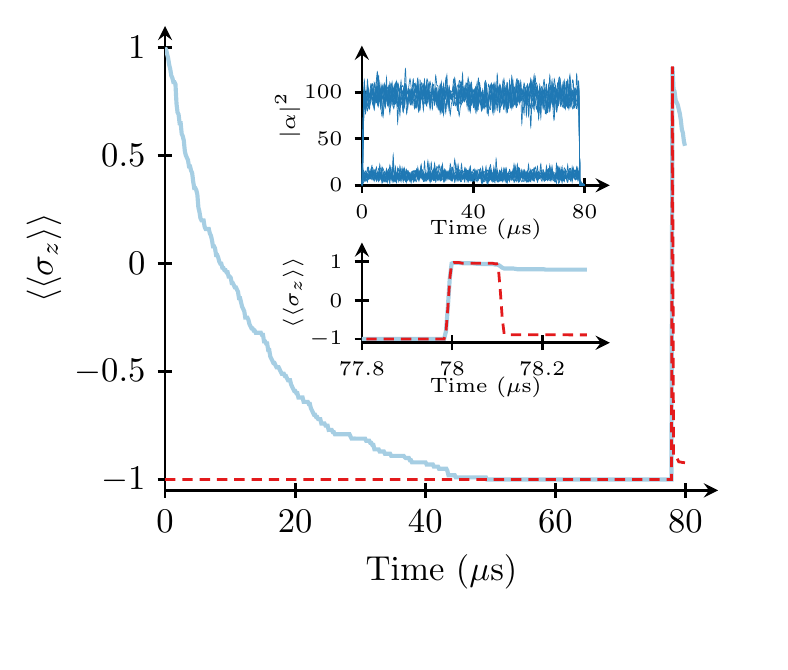}
\caption{(Color online) Demonstration of the memory recovery scheme with parameters described in the text. The main figure shows the mean value of the qubit operator $\sigma_z$ averaged over 200 trajectories for each. The solid light blue line assumes $\ket{1}$ as the initial state, while the dashed red line starts in $\ket{0}$. The $\pi$-pulse is applied after 78 $\mu$s, where we obtain a fidelity with the original state of 98.6\% for $\ket{1}$ and 96.0\% for $\ket{0}$. The upper inset shows Kerr-resonator excitations for 20 randomly chosen realizations and we see two clearly distinct stable states. The lower inset displays a zoom of the region around the recovery of the state.} \label{fig:memory}
\end{figure}

\subsection{Qubit readout and memory}
\label{sec:qbreadmem}
To demonstrate that the classical memory state can be stored and reused in a feedback scheme at a later point in time, we will bring back the measured qubit state after a very long time. This scheme relies upon qubit decay into the ground state, but in practice qubits within a 3d-cavity tend to have a residual thermal excited-state population beyond what is expected from the cryogenic temperature \cite{PhysRevLett.114.240501}. This analysis is thus an idealisation to demonstrate the memory properties of the Kerr-resonator. We now apply a very long $T_{wait}$, and just before applying the tuning sequence, we apply a $\pi$-pulse to bring the qubit into the excited state. The final steps are then the same as in the state-preparation scheme. We have chosen $T_{wait}$, such that we start the tuning after 78 $\mu$s. In Fig. \ref{fig:memory} we illustrate the performance of this memory scheme. From 400 trajectories, we obtain a fidelity for reading out $\ket{1}$ and bringing the qubit back into $\ket{1}$ of 98.6\%, consistent with the state preparation scheme for the excited state. Similarly for $\ket{0}$ we find a fidelity of 96.0\%. The memory scheme is limited by the imperfect $\pi$-pulses as discussed in the previous subsection, thus we have a smaller fidelity for the $\ket{0}$ state. There is also a small (${<}1\%$) remnant excitation from the qubit after $T_{wait}$.

\section{Comparison with other approaches to closed-loop qubit control}

As shown by the numerical simulations, the setup in Fig. \ref{fig:circuit} implements a feedback loop, but in general there is not a single correct way to implement closed-loop control of superconducting microwave qubits, and all-cryogenic and inter-temperature closed loop feedback systems are likely to be both applied in a more mature technology. Thus, it is important to clarify when all-cryogenic control might be naturally competitive. Room temperature measurement-based control \cite{campagne2013persistent, riste2012feedback, riste2013deterministic, vijay2012stabilizing} benefits from mature, flexible, and commercially-available hardware and is easier to debug. But, as already mentioned, room temperature-based feedback control requires interfacing very different and remote technologies.  As such, amplifiers must be state of the art in terms of simultaneous low-noise, high bandwidth, and gain, and gross energy inefficiency is unavoidable between qubits operating at fW power scales and Watt-scale room temperature equipment. In contrast, the components of our proposed scheme are proximate in a cold, low-noise environment, so exemplary amplification is less critical.  Moreover, the qubit and controller feature ``only'' a factor of $\sim10^3$ difference in operating power, and further reduction in operation power is conceivable \cite{andersen2015circuit}. A general concern with cryogenic classical control is the heat dissipation fundamental to digital computation, but in the scheme here, dissipation occurs off-chip, as remote from the sensitive qubit and Kerr-resonator as necessary (unlike, for example, rapid single flux quantum controllers \cite{likharev1991rsfq, chen1999rapid}).

There is also no single correct approach to all-cryogenic feedback control of superconducting qubits.  Often, these schemes are lumped together under the category of ``autonomous control", but there are important distinctions within this category as well. The most popular autonomous approaches today in superconducting microwave systems are continuous pumping schemes, inspired by quantum optics techniques \cite{geerlings2013demonstrating, PhysRevLett.109.183602, shankar2013autonomously, holland2015single}. Here, microwave transitions in a single component (usually a circuit QED qubit or qubits) are designed so that a continuous microwave pump drives the qubit(s) into a desired state.  These schemes are often understood in terms of an internal feedback loop in which a cavity and the qubit states respond to each other. We distinguish this from our approach in that the Kerr-resonator controller may be separated from the qubit without a fundamental change to the qubit's operation. In contrast, remove the cavity from a circuit QED qubit and it becomes a very different device. As a consequence, our qubit and Kerr-resonator controller may be optimized with greater independence.  Also, as the Kerr-resonator's role is well-compartmentalized to a) measuring a digital microwave signal and b) performing feedback with conditional filtering of a microwave pulse, it is easy to see how this concept may apply more generally \cite{PhysRevLett.105.040502} in superconducting microwave systems (See also discussion in Sec. \ref{sec:discout}).  In contrast, strongly coupling a general quantum circuit to a cavity leads to a highly perturbed energy spectrum and it is much less clear how to design such systems for general continuous pumping schemes.

\begin{figure*}
\includegraphics[width=0.75\linewidth]{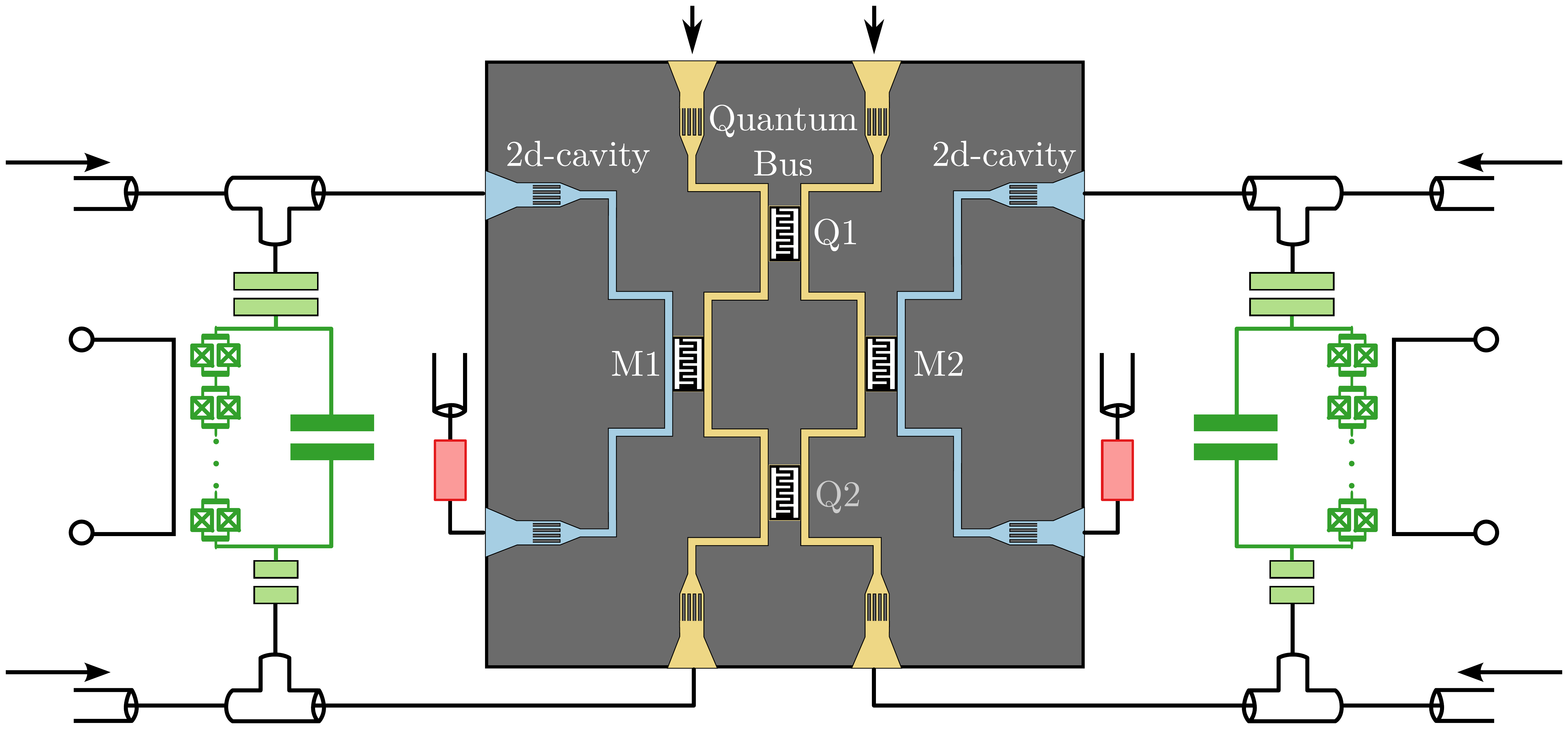}
\caption{(Color online) Schematics to stabilize a Bell-state. The circuit components are the same as in Fig. \ref{fig:circuit}, but instead of 3d-cavities, we use waveguides on a 2d-chip. The state of interest is encoded into the data qubits, Q1 and Q2, while M1 and M2 are measurement qubits and they serve the same purpose as the qubit in Fig. \ref{fig:circuit}. To perform the stabilization of Q1 and Q2, we apply quantum gates between these and M1 and M2. This can be achieved through the two quantum buses (in yellow). Control channels from the Kerr-resonators to each quantum bus allow the feedback to correct errors in the data qubits.} \label{fig:outlook}
\end{figure*}

\section{Discussion and outlook}
\label{sec:discout}
We have proposed a new type of feedback scheme that aims at closing the feedback loop within the cryogenic environment of circuit QED experiments. To achieve this we connect a Kerr-resonator to a 3d-cavity with a qubit. The goal is to control the qubit without having any knowledge of the qubit state outside the cryostat.

As an application of this type of feedback we have simulated three applications. At first we showed that we can prepare the qubit in the excited state with an average fidelity of 97.1\%. Including qubit decay we showed that we can continuously stabilize the excited state at an average fidelity of 95.4\%. Finally, we showed that the scheme also works as a readout with inbuilt classical memory that can bring back the measured qubit state at a much later time with an average fidelity of 97.3\%. The simulations were done using a quantum trajectory approach that includes fluctuations in the Kerr-resonator and 3d-cavity fields and with experimentally realistic parameters.

The main limitation for the schemes is imperfect $\pi$-pulses that suffer from the stochastic perturbations by the Kerr-resonator field and finite readout fidelity. Moreover, the stabilization scheme suffers from the finite lifetime of the qubit, but improving the qubit lifetime and employing Purcell-filter techniques will improve the fidelities significantly.

Although the intention of this work is to accomplish qubit measurement, feedback, and classical memory without extracting information about the qubit state from the cryogenic environment, we must have the option to extract that information to experimentally verify that the concept works as intended (e.g. for tune-up and debugging). In particular, we must verify that after measurement by the Kerr resonator, the qubit state and Kerr resonator state are properly correlated. This can be established with a minimum of added complexity by interrupting the protocol in Fig.~4 immediately after the first $T_\Delta$ period. At that time, we can measure the state of the qubit by applying a pulse into the $\alpha_{in}$ port, near resonance with the 3d-cavity and of appropriate amplitude and duration to constitute a so-called high power readout \cite{reed2010high}. By monitoring the reflected pulse phase, even with an inefficient microwave amplifier, we achieve a high-fidelity, but destructive readout of the qubit state. As the qubit cavity system has a much stronger non-linearity than the Kerr resonator, and the Kerr resonator is far detuned from the qubit cavity at this moment in the protocol, we expect that the high power readout can be accomplished without altering or being altered by the state of the Kerr resonator. Subsequent to the high power qubit readout, we can easily determine the state of the Kerr resonator by exciting it with a very weak tone added to the $\alpha_d$ port. The phase shift of this weak tone, monitored again through the $\alpha_{in}$ port, will depend on the state of the Kerr resonator. Because this second measurement detects the stable state of the Kerr resonator, it can be performed long after the qubit readout when all fields from the high-power readout have decayed away. It should likewise be independent of the detected qubit state. In this manner we can measure the classical correlation between the qubit and Kerr resonator state that exist at the time when the high-power qubit readout is performed.

Finally it will be natural to extend the system to include more qubits or more resonators. To mitigate the need for qubit dissipation for the reinitialization protocol described in Sec \ref{sec:qbreadmem} and to suppress residual qubit excitation, a second Kerr-resonator can be used to prepare the qubit in the ground state, just before the memory state, stored in the first Kerr-resonator, is fed back to the qubit.

As already discussed, the protocols presented in this work naturally extend to multiple qubits. Controlling many qubits with closed feedback outside of the cryogenic environment poses a significant challenge \cite{asaad2015independent} since each qubit-readout requires its own signal analyzer and a corresponding signal generator to apply the desired feedback. For systems with 1000 qubits or more, such a setup might be infeasible even with state-of-the-art multi-channel analyzers. Therefore, even initialization of all qubits may seem an overwhelming task. In contrast, the all-cryogenic state-preparation presented in this work is inherently parallel, as each qubit has it own memory that holds the measured qubit state and autonomously controls the input state. The only additional requirement to prepare any number of qubit in, e.g., the ground state is that enough power is provided and appropriate power dividers are installed to drive all qubits sytems.

A more elaborate extension allows stabilization of e.g. a Bell-state of two qubits against phase-flip and bit-flip errors using a quantum error correcting code. A possible setup for this employs two data qubits in which the Bell-state is encoded and two additional measurement qubits to detect the errors. Performing two-qubit gates between the data and the measurement qubits followed by a readout of the two measurement qubits will detect any errors in the encoded state \cite{PhysRevA.86.032324, corcoles2015demonstration}. A subsequent feedback on the data qubit will stabilize the state. In Fig. \ref{fig:outlook} we have sketched how such a scheme might be implemented using a 2d-chip with multiple qubits \cite{riste2015detecting}. Two units, each consisting of a Kerr-resonator, a 2d-cavity and a measurement qubit, allow us to perform the desired two-qubit gates and the additional coupling between the quantum buses and the Kerr-resonators will be sufficient to effectively stabilize the desired Bell-state in a completely autonomous manner. Extending this scheme to a seven qubit code is in principle possible and would allow autonomous stabilization of any logical qubit state with a limited amount of external control lines.

The Bell-state stabilization represents a minimal implementation of the so-called surface code \cite{PhysRevA.86.032324, Dennis02topological}. In general, the surface code can be extended to any number of qubits, and the outcome of the measurements identifies whether an error occurred or not at a given point in the code. If an error is found, it is not necessary to correct it immediately, but errors can be tracked and corrected at the end of the full protocol \cite{PhysRevA.86.032324}. All-cryogenics control may implement Kerr-resonator memories in a full-scale quantum error correction codes and, thus, may become vital in future quantum information technologies.

\section*{Acknowledgements}
CKA and KM acknowledge support from the Villum Foundation Center of Excellence, QUSCOPE, and CKA further acknowledges support from the Danish Ministry of Higher Education and Science. JK, KWL and BJC acknowledge support from the ARO under contract W911NF-14-1-0079.

\begin{appendix}
\section*{APPENDIX: Comparison of mean field and fully quantum description}
\label{sec:appa}
%
%

\begin{figure}[bt]
\includegraphics[width=\linewidth]{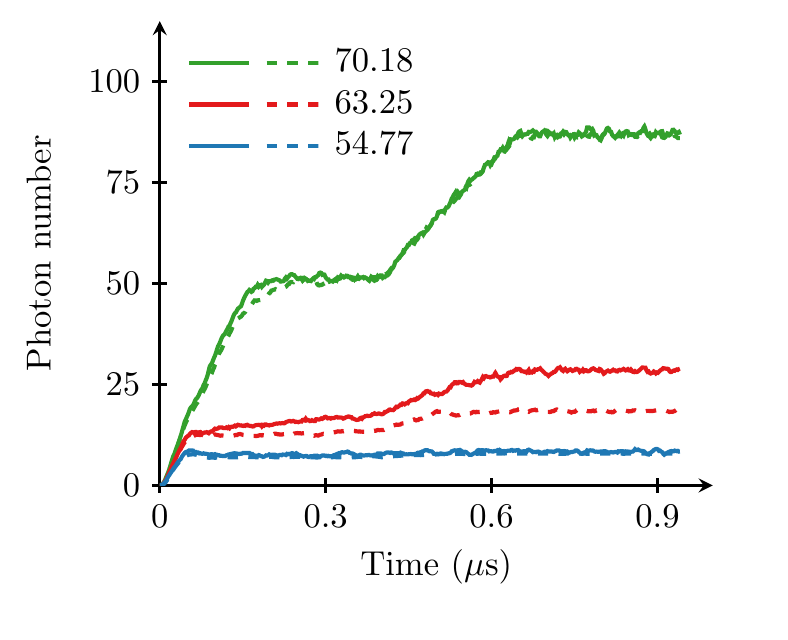}
\caption{A simulation of the Kerr-resonator with the similar parameters as in the main text, but with $\kappa_b$ = 0. The solid lines are the semi-classical approach described in Sec. \ref{sec:full}, while the dashed lines are the same but using the Monte-Carlo wavefunctions of the Appendix. In both types of simulation we have averaged over 100 trajectories. The legend indicate the value of drive expressed as $\alpha_d / \sqrt{\kappa_d}$.} \label{fig:semicheck}
\end{figure}

To compare the semi-classical approach of Eq. \eqref{eq:alpha} with the fluctuations of Eq. \eqref{eq:xia}, we simplify the model, such that we only have the Kerr-resonator, that is, we set $\kappa_b = 0$ in Eq. \eqref{eq:alpha}. To describe the Kerr-resonator in a fully quantum calculation, we choose to calculate quantum trajectories using Monte-Carlo wavefunctions (MCWF), such that the unnormalized wavefunction is given by the non-unitary Schrödinger equation,
\begin{appendixalign}
\frac{d}{dt} \ket{\psi} = -\frac{i}{\hbar}H \ket{\psi} - \frac{\kappa_d + \kappa_a}{2} a\dag a \ket{\psi},
\end{appendixalign}
with the Hamiltonian
\begin{appendixalign}
H = \hbar\Delta_a a\dag a + \hbar\frac{K}{2} a\dag a a\dag a + i\hbar \sqrt{\kappa_d} \alpha_d (a\dag - a).
\end{appendixalign}
Similar to the qubit description in main article, we evolve this wavefunction until a the norm becomes smaller than a given random number, $R$, drawn uniformly between 0 and 1. We then apply the jump operator $a$ to the wavefunction, we then renormalize and draw a new $R$. On average this reproduces a full master equation description and it is expected to produce the same behaviour as the semi-classical description. Using the unnormalized wavefunction we can calculate any mean value as \mbox{$\exv{X} = 1/M \sum_m \bra{\psi_m} X \ket{\psi_m} / \braket{\psi_m}{\psi_m}$}, with $M$ being the total number of trajectories, $\ket{\psi_m}$.

In Fig. \ref{fig:semicheck} we see examples of a MCWF simulation and an equivalent semi-classical simulation. We these simulation we see that the strong driving and the weak driving matches perfectly. We even see that the memory effect, which might seem like a classical artefact of the semi-classical approach is reproduced in the quantum simulation. For the middle driving strength we do, however, notice a discrepancy between the two methods. We ascribe this seemingly different behaviour to a difference in the exact bifurcation point. Looking at the individual trajectories we see a very similar behaviour between the two methods, but for the full quantum approach we find 10\% of the trajectories latching into a high photon number state, while this number is 22\% for the semi-classical approach. With these results in hand we can therefore conclude that the the behaviour of the semi-classical approach in the main paper will produce results similar to a full quantum implementation, but we expect that the required drive strength must be re-calibrated.

\end{appendix}

\bibliography{bt}

\end{document}